\documentstyle [12pt,dina4,psfig,amssymb]{article}
\begin{document}
\parindent=0cm
\parskip=1.5mm

\def\bi{\begin{list}{$\bullet$}{\parsep=0.5\baselineskip
\topsep=\parsep \itemsep=0pt}}
\def\ei{\end{list}}

\def\phi{\varphi}
\def\-{{\bf --}}
\def\vm{v_{max}}
\newcommand{\s}{\sigma}
\newcommand{\la}{\lambda}
\newcommand{\eps}{\varepsilon}
\newcommand{\al}{\alpha}
\newcommand{\nn}{{\cal N}}

\newcommand{\be}{\begin{equation}}
\newcommand{\ee}{\end{equation}}
\newcommand{\bea}{\begin{eqnarray}}
\newcommand{\eea}{\end{eqnarray}}

\newcommand{\nonu}{\nonumber\\}

\renewcommand{\thefootnote}{\fnsymbol{footnote}}

\newcommand{\tP}{\tilde P_1}
\newcommand{\bP}{\bar P_1}

\parindent=0cm
\begin{center}
  {\LARGE\bf Traffic flow models with}
\end{center}
\begin{center}
  {\LARGE\bf 'slow-to-start' rules}
\end{center}
\vskip1.8cm \renewcommand{\thefootnote}{\fnsymbol{footnote}}
\setcounter{footnote}{1}
\begin{center}
  {\Large Andreas Schadschneider$^{1}$ and Michael Schreckenberg$^{2}$}
\end{center}
\vskip1.3cm
\begin{center}
  $^{1}$ Institut f\"ur Theoretische Physik\\ Universit\"at zu K\"oln\\ 
  D--50937 K\"oln, Germany\\
  email: {\tt as@thp.uni-koeln.de}
\end{center}
\begin{center}
  $^{2}$ Theoretische Physik/FB 10\\ 
  Gerhard-Mercator-Universit\"at Duisburg\\ 
  D--47048 Duisburg, Germany\\
  email: {\tt schreck@uni-duisburg.de} 
\end{center}
\begin{center}
\today
\end{center}
\vskip1.3cm \vskip2cm {\large \bf Abstract}\\[0.2cm]
We investigate two models for traffic flow with 
modified acceleration ('slow-to-start') rules. Even in the 
simplest case $\vm=1$ these rules break the 'particle-hole` symmetry 
of the model.
We determine the fundamental diagram (flow-density relationship)
using the so-called car-oriented mean-field approach (COMF) which
yields the exact solution of the basic model with $\vm=1$. 
Here we find that this is no longer true for the models with
modified acceleration rules, but the results are still in good
agreement with simulations. We also compare the effects of the
two different slow-to-start rules and discuss their relevance for
real traffic. In addition, in one of these models we find a new 
phase transition to a completely jammed state.\\[1cm]
{\bf Keywords:} Cellular automata, traffic flow, improved mean-field theories
\vfill
\pagebreak
\section{Introduction}

The investigation of traffic flow has attracted a lot of interest
from physicists in the last few years. In addition to more traditional
methods borrowed e.g.\ from hydrodynamics, ideas from statistical
physics have become more and more important. 
A comprehensive review of the different approaches can be found in 
\cite{Jue}.

Here we use a cellular automaton (CA) \cite{wolfram} which has been
shown to reproduce at least qualitatively the features of 
real traffic \cite{ns,ssni}. The basic idea is to formulate a model
in discrete space and time. The road is divided into cells which can
either be empty or occupied by exactly one car. The cars are characterized
by an internal parameter $v=0,1,\ldots,\vm$ which corresponds to their 
momentary velocity. Starting from a given configuration at a time $t$ the
configuration at the next time step $t+1$ can be obtained by application
of four rather simple rules which are applied to all cars in parallel: 
acceleration, slowing down due to other cars, randomization and the 
actual car motion. 

The stochastic nature of the model which enters through the randomization
process seems to be important for a realistic description of traffic.
It is quite surprising that such a simple rule can mimic several effects 
which seem to be important for the behaviour of real traffic, e.g.\
overreactions at braking or fluctuations in driving.

Due to their inherent discrete nature CA are ideal for 
large-scale computer simulations. On the other hand, these models can not be
described easily using analytic approaches. 
In \cite{ss} we solved the special case of $\vm=1$ exactly.  We calculated
the so-called fundamental diagram, i.e.\ the relation between flow (current)
and the density $c$ of cars. The case $\vm=1$ is special since the model
is particle-hole symmetric and therefore the fundamental diagram is
also symmetric with respect to $c=1/2$ where also the flow takes its
maximum value. From measurements on real traffic one knows, however,
that the maximum flow is shifted to much lower densities. This fact
is reproduced by the models with $\vm>1$, but unfortunately these models
are no longer exactly solvable.

In \cite{benja,taka} two models with slightly modified acceleration
rules have been introduced. One effect of these new rules is a breaking
of the particle-hole symmetry in the case $\vm=1$. In this paper we
want to address the question whether these modified versions are still
exactly solvable or if the exact solvability is somehow connected to
the existence of the particle-hole symmetry. Another interesting point
is a comparision of the two rules. Both are designed to mimic the
same effects observed in real traffic, but use different ways to implement
these in the rules of the CA. We therefore want to compare the effects of the 
two different slow-to-start rules and discuss their relevance for modelling
real traffic.

\section{The modified models}

In the following we will always assume $\vm=1$. The fundamental
diagram of the basic model is symmetric (maximum flow at density $c_{max}=1/2$)
due to a particle-hole symmetry \cite{ss}. In order to get a more realistic
asymmetric fundamental diagram even for $\vm =1$ modified models with
new acceleration rules have been introduced \cite{benja,taka}.

In the Benjamin-Johnson-Hui (BJH) model \cite{benja} cars which had to
brake due to the next car ahead will move on the next opportunity only
with probability $1-p_s$. Each cell can be in one of four states.
Either it is empty (state `e') or it is occupied by a car with
velocity $v=1$ (state `1'), a car with velocity $v=0$ which is static
due to the action of the randomization step or the slow-to-start rule
(state `s'), or a car with velocity $v=0$ which is standing due to
blocking by a car just in front of it (state `b').

\bi

\item[1)] {\bf Acceleration:} 
All cars are assigned a velocity of `1'.

\item[2)] {\bf Slow-to-start rule:}
All cars that are legitimate candidates decelerate to `s' with probability
$p_s$.

\item[3)] {\bf Blockage:} 
All cars which have no free cell in front and are thus blocked from
moving change to state `b'.

\item[4)] {\bf Randomization:}
All cars still in state `1' decelerate to state `s' with probability $p$.

\item[5)] {\bf Car motion:}
All cars in state `1' move one cell ahead.

\ei

In the Takayasu (T$^2$) model \cite{taka} a standing car will move only 
if there are at least two free cells in front of it. More generally it 
will move only with probability $1-p_t$ if there is just one free cell 
ahead. Here every cell is one of three states, i.e.\ empty, or occupied
by a car with velocity $v=0$ or $v=1$.

\bi

\item[1)] {\bf Acceleration:} 
Standing cars accelerate to velocity $v=1$ if there are at least
two empty cells in front.

\item[2)] {\bf Slow-to-start rule:}
Standing cars with just one free cell in front accelerate only with 
probability $q_t=1-p_t$ to velocity $v=1$.

\item[3)] {\bf Slowing down (due to other cars):} 
If the distance $d$ to the next car ahead is not larger than $v$ ($d \le v$)
the speed is reduced to $d-1$ [$v \stackrel{d \le v}{\to}d-1$].

\item[4)] {\bf Randomization:}
With probability $p$, the velocity of a vehicle
(if greater than zero) is decreased by one [$v\stackrel{p}{\to}v-1$]. 

\item[5)] {\bf Car motion:}
Each vehicle is advanced $v$ cells.

\ei

The idea behind the modified rules of the BJH and T$^2$ models is to
mimic the delay of a car in restarting, i.e.\ due to a slow pick-up of
engine or loss of the driver's attention. Both modifications break 
the particle-hole symmetry and should yield fundamental diagrams with a 
maximum at a density $c_{max}<1/2$.

In order to investigate the effects of the modified rules analytically
we apply the so-called ``car-oriented mean-field theory'' (COMF)
\cite{comf}. This approach turned out to yield the exact solution for
the standard rules ($p_s=0$, $p_t=0$) \cite{comf}. It seems therefore
worthwhile to investigate whether one also obtains the exact solution
for the modified models without particle-hole symmetry.

\section{COMF for BJH model}

We first investigate the effects of the BJH version \cite{benja} 
of the slow-to-start rule.

We denote the probability to find at time $t$ (exactly) $n$ empty
cells in front of a vehicle by $P_n(t)$. The density of cars which
which obey the slow-to-start rule is denoted by $\tilde P_1$. As in
\cite{ss,ssni} we change the order of the update steps to 2-3-4-1.
This change has to be taken into account when calculating the flow
$f(c,p,p_s)$. It has the advantage that after step 1 there are no cars
with velocity 0, i.e.\ all cars have velocity 1.

The time evolution of the probabilities $P_n$ can conveniently be
expressed through the probability $g(t)$ ($\bar g(t)=1-g(t)$) that
a car moves (does not move) in the next timestep. These probabilities
are given by:
\bea
g(t)&=&q\sum_{n\geq 1} P_n(t) + q_sq\tP(t) 
= q[1-P_0(t)]-p_sq\tP(t),
\label{gdef1}
\eea
where we have used the normalization
\be
\sum_{n\geq 0} P_n(t) + \tP(t)= 1.
\label{norm1}
\ee
The probabilities can be related to the density $c=N/L$ of cars. Since
each car which has the distance $n$ to the next one in front of him
'occupies' $n+1$ cells we have the following relation:
\be
\sum_{n\geq 0}(n+1)P_n(t) + 2\tP (t) = \frac{1}{c}\ .
\label{dens1}
\ee

As described in \cite{comf} we then obtain the time evolution of the 
probabilities as ($q=1-p$, $q_s=1-p_s$)
\bea
P_0(t+1)&=& \bar g(t)[P_0(t)+qP_1(t)+q_sq\tP(t)]               \label{evol1a}\\
\tP(t+1)&=& g(t) P_0(t)                                        \label{evol1b}\\
P_1(t+1)&=& \bar g(t)[pP_1(t)+(p_s+q_sp)\tP(t)+qP_2(t)]
+q g(t)[P_1(t)+q_s\tP(t)]                                      \label{evol1c}\\
P_2(t+1)&=& \bar g(t)[pP_2(t)+q P_3(t)]
+ g(t)[pP_1(t)+(p_s+q_sp)\tP(t)+qP_2(t)]                       \label{evol1d}\\
P_n(t+1)&=& \bar g(t)[pP_{n}(t)+qP_{n+1}(t)]+g(t)[pP_{n-1}(t)+qP_n(t)]
\quad (n\geq 3)                                                \label{evol1e}
\eea
Here we are mainly interested in the stationary state ($t\to\infty$) with
$\lim_{t\to\infty} P_n(t)=P_n$. 
In \cite{comf} we used generating functions to solve a similar set of
equations ($p_S=0$). Here we adopt a more direct method since it is easy to see
that (\ref{evol1e}) is solved by the Ansatz
\bea
P_n&=&\nn z^{n}, \qquad \qquad \qquad (n\geq 2) \nonu
z&=&\frac{pg}{q\bar g}\, ,
\label{ansatz1}
\eea
where $\nn$ is a normalization.

Using (\ref{evol1a}-\ref{evol1d}) we can express all probabilities through
$P_0$:
\bea
\tP &=& \frac{qP_0(1-P_0)}{1+p_sqP_0}, \\ 
g&=&\frac{\tP}{P_0}=\frac{q(1-P_0)}{1+p_sqP_0},\\
P_1 &=& \frac{zP_0}{p} - q_s\tP,\\
\nn&=&(1-P_0-P_1-\tP)\frac{1-z}{z^2}.
\eea

The only remaining free parameter $P_0$ can now be related to the density 
$c$ by using (\ref{dens1}). $P_0(c)$ 
is given as the root in the interval $(0,1)$ of the cubic equation
\be
cp_s^2q^2P_0^3+q[qp_s^2(1-2c)+p_s(1+c)+c]P_0^2+[qp_s(1-3c)-2qc+1]P_0-pc=0.
\ee

The flow is given by (with $q=1-p$)
\be
f(c,p,p_s)=cg=c\cdot\frac{q(1-P_0)}{1+p_sqP_0}.
\label{flow1}
\ee
Results for the fundamental diagram and a comparison with computer
simulations are shown in Figs.\ \ref{BJHp} and \ref{BJHps}. Obviously, the
COMF is no longer exact for $p_s>0$, but it is still an excellent 
approximation, especially for small $p_s$. For fixed $p$ (see Fig.\ \ref{BJHp})
the flow $f(c,p,p_s)$ is only reduced slightly for $p_s>0$. In addition,
the density $c_{max}$ of maximum flow is shifted towards smaller
densities with increasing $p_s$, i.e.\ the fundamental diagram is no 
longer symmetric. This reflects the absence of the particle-hole symmetry.
There is no other qualitative change of the fundamental 
diagram, in contrast to the T$^2$ model (see next section).


\section{COMF for the T$^2$ model}

The probability $g(t)$ ($\bar g(t)=1-g(t)$) that a car moves (does not move) 
in the next timestep is now given by:
\bea
g(t)&=&q\sum_{n\geq 1} P_n(t) + q_tq\bP(t) 
= q[1-P_0(t)]-p_tq\bP(t),
\label{gdef2}
\eea
We obtain the time evolution of the probabilities as ($q=1-p$, $q_t=1-p_t$)
\bea
P_0(t+1)&=& \bar g(t)[P_0(t)+qP_1(t)+q_tq\bP(t)]               \label{evot1a}\\
\bP(t+1)&=& \bar g(t) (p_t+q_tp)\bP(t)+g(t) P_0(t)             \label{evot1b}\\
P_1(t+1)&=& \bar g(t)[pP_1(t)+qP_2(t)]+qg(t)[P_1(t)+q_t\bP(t)] \label{evot1c}\\
P_2(t+1)&=& \bar g(t)[pP_2(t)+q P_3(t)]
+ g(t)[pP_1(t)+(p_t+q_tp)\bP(t)+qP_2(t)]                       \label{evot1d}\\
P_n(t+1)&=& \bar g(t)[pP_{n}(t)+qP_{n+1}(t)]+g(t)[pP_{n-1}(t)+qP_n(t)]
\quad (n\geq 3)                                                \label{evot1e}
\eea
Here we are mainly interested in the stationary state ($t\to\infty$) with
$\lim_{t\to\infty} P_n(t)=P_n$. In this case the equations 
(\ref{evot1e}) -- which are identical with the equations (\ref{evol1e})
since the modified rules do not affect distances larger than 1 --
are again solved by the Ansatz 
\bea
P_n&=&\nn z^{n}, \qquad \qquad \qquad (n\geq 2) \nonu
z&=&\frac{pg}{q\bar g}\, ,
\label{ansatz2}
\eea
where $\nn$ is a normalization.
Expressing all quantities through $P_0$ and $\bP$ we find
\bea
P_1&=&\frac{g}{q\bar g}\bP = \frac{z}{p}\bP ,\\
g&=&q(1-P_0)-p_tq\bP,\\
\nn &=&(1-P_0-P_1-\bP)\frac{1-z}{z^2}.
\eea
$P_0$ and $\bP$ can be related through
\be
p_tx\bP^2 - (1+pq_t(1-P_0))\bP+(1-P_0)P_0 = 0
\ee
which yields explicitly
\be
\label{wurzel}
\bP = \frac{1}{2p_tx}\left[ 1+pq_t(1-P_0)-\sqrt{(1+pq_t(1-P_0))^2-
4p_tx(1-P_0)P_0}\right]
\ee
where we have used the abbreviation $x=p_t+q_tp=1-qq_t$.

The relation with the density is 
\be
y(c-1)+(p+(1+q)y)c(1-P_0)-(1+p-py)c\bP = 0
\label{tdens}
\ee
with $y=P_0+p_t\bP$.
After solving (\ref{tdens}) for $P_0(c)$, $\bP(c)$  the flow can 
be obtained using $f(c,p,p_t)=cg$. In Figs.\ \ref{T2p} and \ref{T2ps} we show 
the results of COMF and computer simulations. Again the COMF is no longer
exact for $p_t>0$, but it is still a good approximation. For $p_t \ll 1$
the situation is somewhat similar to the BJH model, i.e.\ the flow is
reduced slightly and the $c_{max}$ is shifted towards smaller densities.
For large $p_t$, however, one finds a qualitative change of the fundamental
diagram. For $p_t \lesssim 1$ the fundamental diagram has an inflection 
point at a density $c>c_{max}$, in contrast to the case $p_t \ll 1$,
where the flow vs.\ density relation is convex.
The parameter $p_t$ therefore controls the curvature of the
fundamental diagram.

This can be seen most clearly for $p_t=1$. Here the flow $f(c,p,p_t)$
vanishes for $c \geq 1/2$. For $c=1/2$ the stationary state consists
of standing cars with exactly one empty site in between them ($0.0.0.0.
\cdots$ where '$0$' denotes a standing car and '.' an empty site). All cars
cannot move due to the slow-to-start rule. For larger densities the stationary
state is essentially of the same form, but larger clusters of standing
cars will appear. In Monte Carlo simulations
starting from a mega-jam the relaxation into this stationary state is
extremly slow, even for 'large' values of $p$, e.g.\ $p=0.5$. Note that the
COMF solution (\ref{ansatz2})-(\ref{tdens}) predicts a phase 
transition to the completely jammed state at $c=2/3$.
There is, however, a second solution\footnote{This solution corresponds
to (\ref{wurzel}) with a '+' in front of the square-root.} of the 
COMF equations with $P_0=1-\bar{P}_1=\frac{2-c}{c}$ and $P_n=0$ for $n\geq 1$,
corresponding to $f(c,p,p_t=1)=0$ for $c\geq 1/2$.

The new phase transition for $p_t=1$ only exists for $p>0$. 
In \cite{taka,fukui} mainly the
deterministic case $p=0$ has been investigated. There the equilibrium
state depends on the initial conditions and therefore the completely
jammed phase for $c\geq 1/2$ could not been found. More detailed results
for the model with $p_t=1$ will be presented in a future publication.

\section{Discussion}

In this paper we have investigated two cellular automata for
traffic flow with modified acceleration rules. For $\vm =1$ these rules
break the particle-hole symmetry of the model. We used the so-called
Car-Oriented Mean-Field theory to obtain an analytic desription of these
models. In contrast to the case of the basic model \cite{comf} 
the COMF does no longer produce an exact solution of the modified versions.

The main effects of the slow-to-start rules are a (small) reduction of the
flow and a shift of the location $c_{max}$ of the maximum flow towards smaller
densities. Note, however, that the reduction in the T$^2$ model is larger
than in the BJH model, especially for large $p_t$. For $p_t=1$ we found a
phase transition to a completely jammed phase with flow $f(c,p,p_t=1)=0$ 
near $c=1/2$. This transition will be investigated in more detail in a 
future publication.

The existence of this transition reflects the most important
difference between the two slow-to-start rules. The BJH rule is of
{\em temporal} nature (the restart probability depends on the number
of attempts) whereas the T$^2$ rule is of {\em spatial} nature (the
restart probability depends on the number of empty cells in front of
the car).

In Figs.\ \ref{tildeP} and \ref{barP} we show the numbers $\tP(c)$ and
$\bP(c)$ of cars affected by the slow-to-start rule in the BJH and T$^2$ 
model, respectively. As expected the modified rules are most effective
for densities slightly larger than the density of maximum flow in the
original model, i.e.\ for $c>1/2$. This effect is most pronounced for
small values of $p$.

Comparing the effects of the two modifications, the T$^2$ model
appears to be more interesting since the effects are larger.
An interesting feature of this model is that the
curvature of the fundamental diagram for densities $c>1/2$ may vary,
depending on the value of $p_t$. The idealized shape of the
fundamental diagram has been discussed controversely in the literature
(see e.g.\ \cite{hall}) and there exists experimental evidence for
fundamental diagrams with and without inflection points. The slow-to-start
rule of T$^2$ offers a simple explanation for these observations.

In the present paper we only investigated the case $\vm=1$. The main 
modification introduced by the slow-to-start rules is a asymmetry between
acceleration and deceleration processes. Since this is already included
in the basic model for $\vm>1$ we do not expect that the modifications
might have a drastic effect for larger velocities. This expectation seems
to be justified as preliminary results from computer simulations show.
For the T$^2$ model, however, the slow-to-start probability $p_t$ still
controls the curvature of the fundamental diagram.\\[0.8cm]
{\bf Acknowledgment:}\\
Part of this work has been performed within the research program of
the SFB 341 (K\"oln--Aachen--J\"ulich). 


{\newpage
\begin{figure}[ht]
   \centerline{\psfig{figure=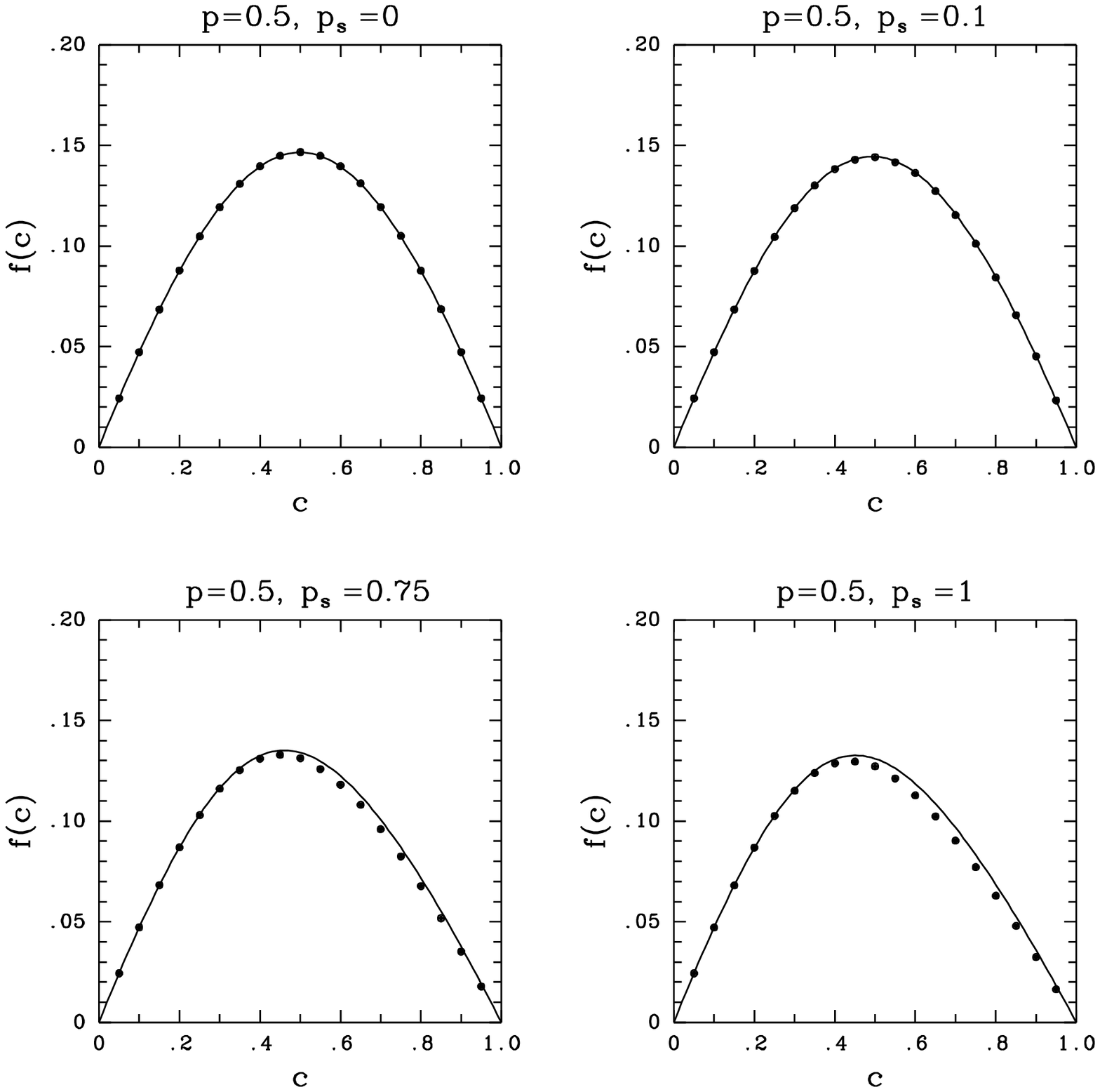,bbllx=30pt,bblly=130pt,bburx=580pt,bbury=685pt,height=12cm}}
 \caption{\protect{Fundamental diagram for the BJH model with $p=0.5$ and
different $p_s$. The full line is the COMF result. For comparison the results 
from computer simulations ($\bullet$) are also shown. }}
\label{BJHp}
\end{figure}

\begin{figure}[ht]
   \centerline{\psfig{figure=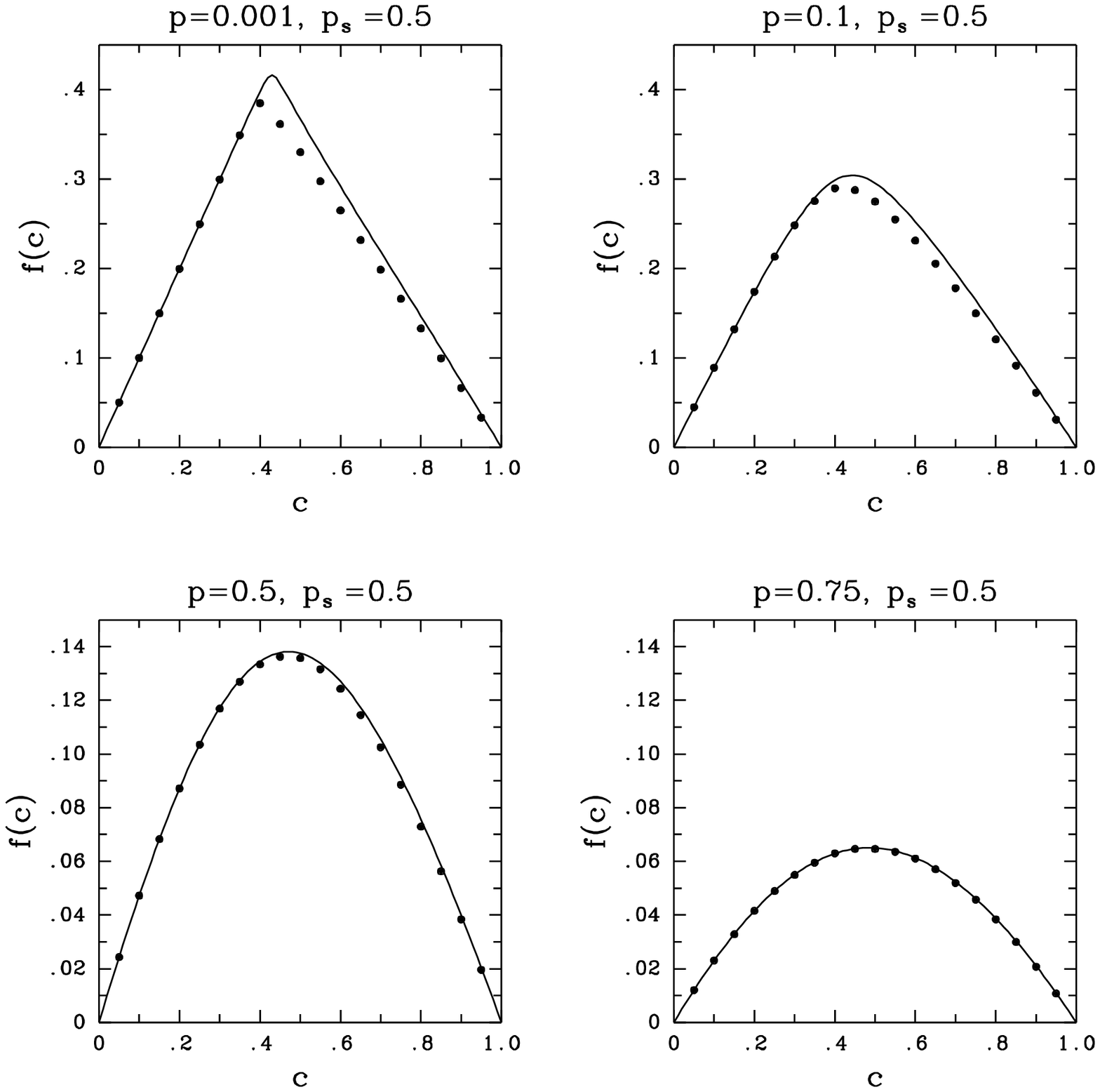,bbllx=30pt,bblly=130pt,bburx=580pt,bbury=685pt,height=12cm}}
 \caption{\protect{Same as Fig.\ \ref{BJHp}, but for $p_s=0.5$ and 
different $p$.}}
\label{BJHps}
\end{figure}

\begin{figure}[ht]
   \centerline{\psfig{figure=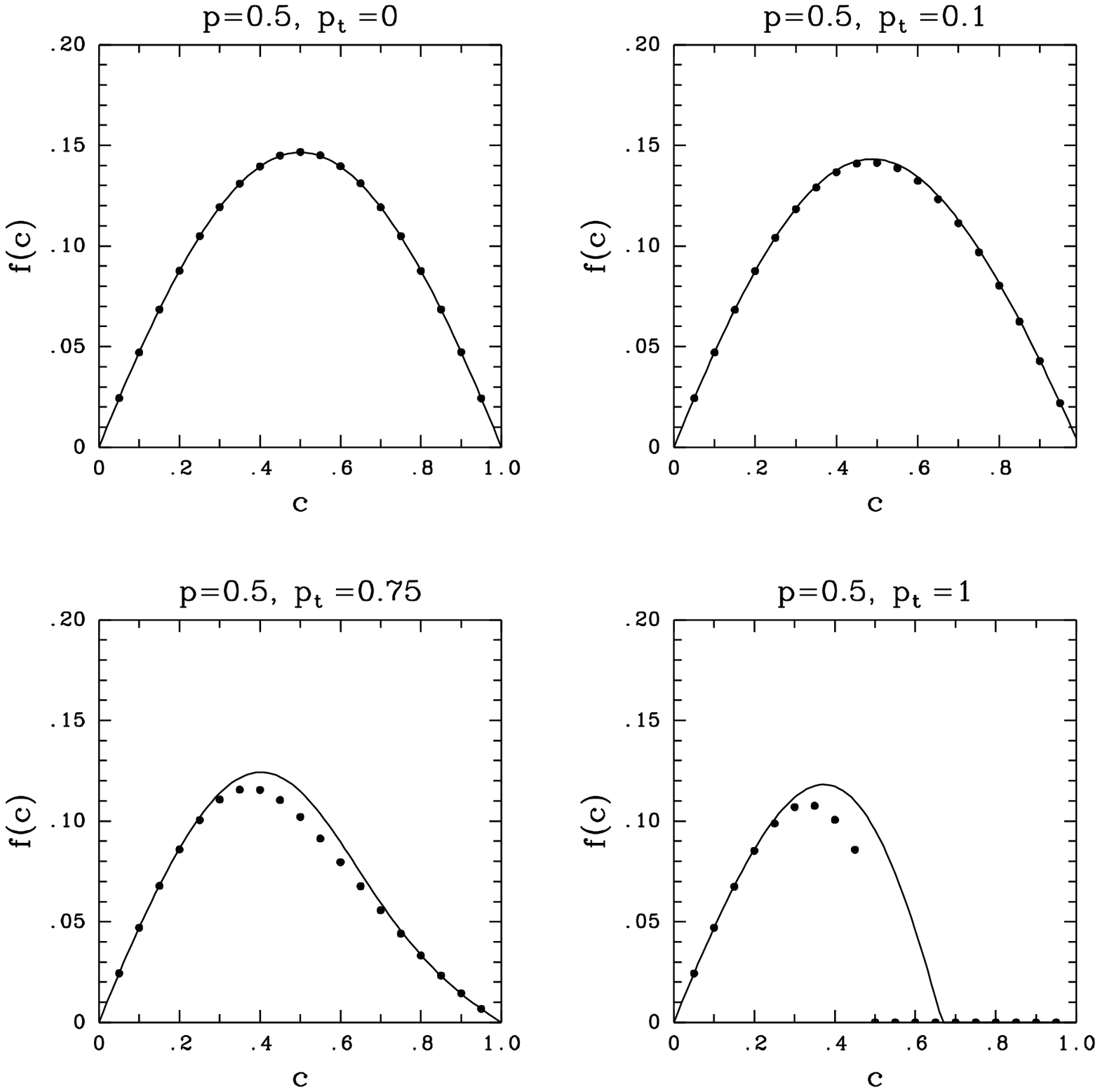,bbllx=30pt,bblly=130pt,bburx=580pt,bbury=685pt,height=12cm}}
 \caption{\protect{Fundamental diagram for the T$^2$ model with $p=0.5$ and
different $p_s$. The full line is the COMF result. For comparison the results 
from computer simulations ($\bullet$) are also shown.}}
\label{T2p}
\end{figure}

\begin{figure}[ht]
   \centerline{\psfig{figure=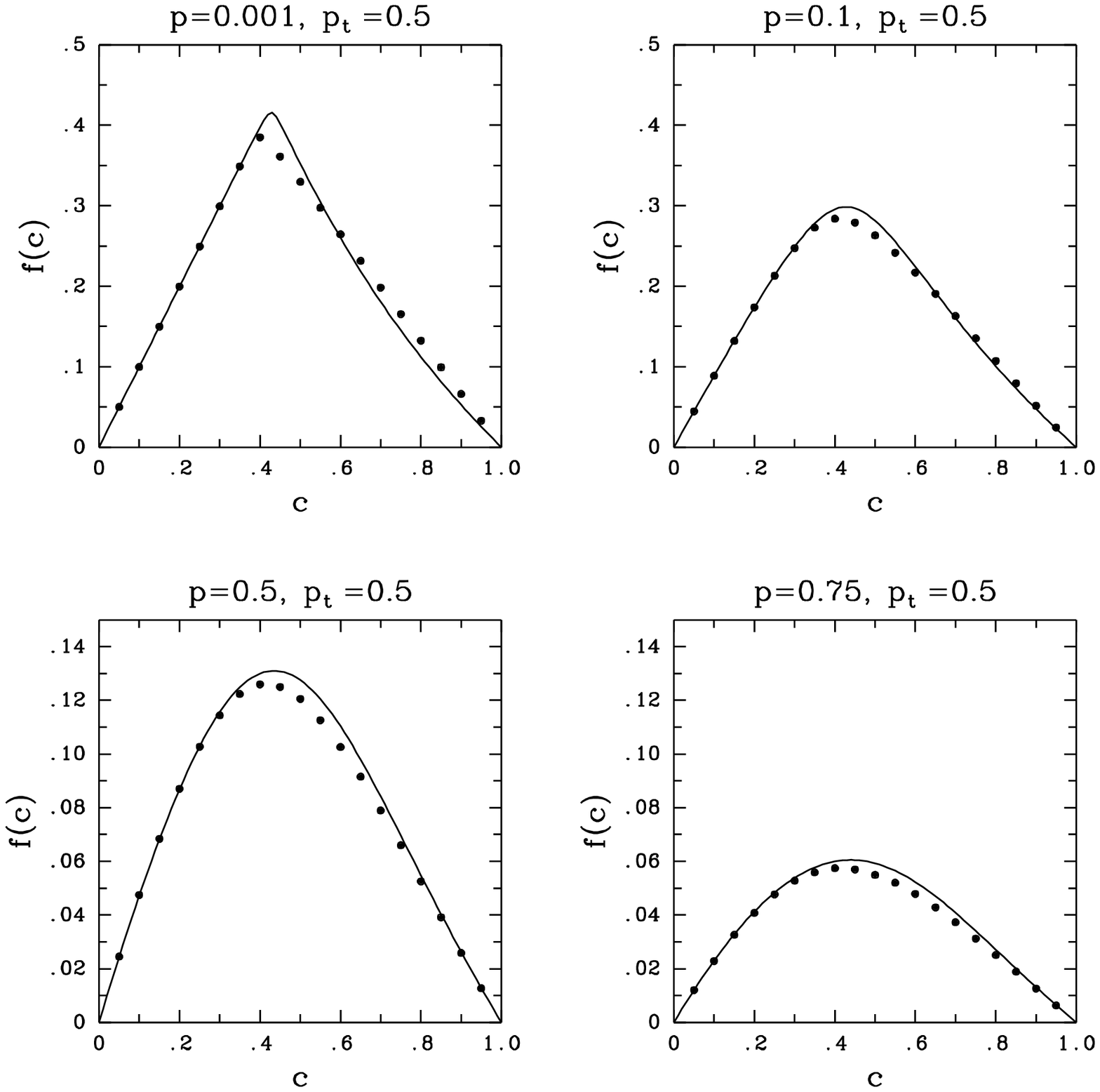,bbllx=30pt,bblly=130pt,bburx=580pt,bbury=685pt,height=12cm}}
 \caption{\protect{Same as Fig.\ \ref{T2p}, but for $p_s=0.5$ and different 
$p$.}}
\label{T2ps}
\end{figure}

\begin{figure}[ht]
   \centerline{\psfig{figure=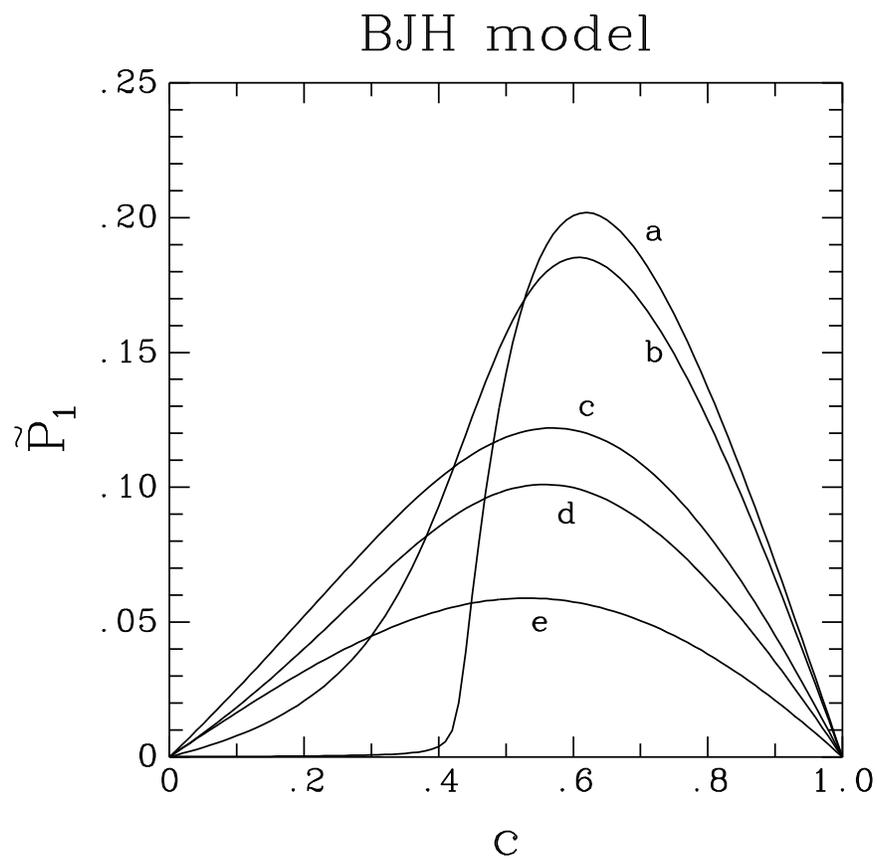,bbllx=70pt,bblly=160pt,bburx=560pt,bbury=680pt,height=12cm}}
 \caption{\protect{$\tP$ for the BJH model for a) $p=0.001$, $p_s=0.5$, 
b) $p=0.1$, $p_s=0.5$, c) $p=0.5$, $p_s=0.1$, d) $p=0.5$, $p_s=1$,  
e) $p=0.75$, $p_s=0.5$.}}
\label{tildeP}
\end{figure}

\begin{figure}[ht]
   \centerline{\psfig{figure=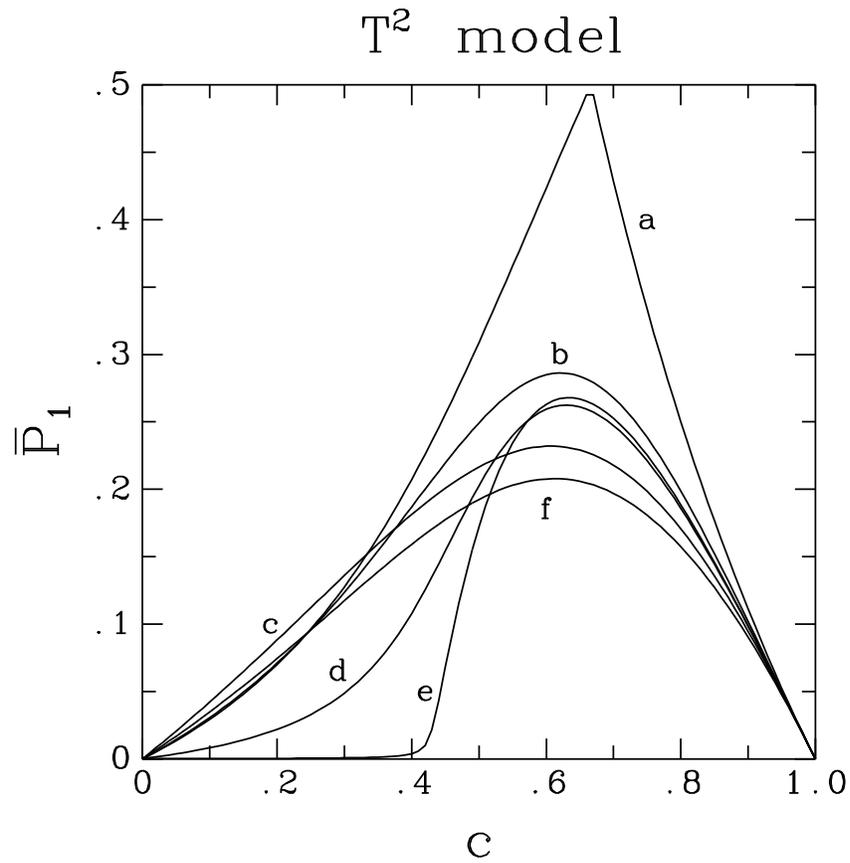,bbllx=70pt,bblly=160pt,bburx=560pt,bbury=680pt,height=12cm}}
 \caption{\protect{$\bP$ for the T$^2$ model for a) $p=0.5$, $p_s=1$, 
b) $p=0.5$, $p_s=0.75$, c) $p=0.75$, $p_s=0.5$, d) $p=0.1$, $p_s=0.5$,  
e) $p=0.001$, $p_s=0.5$.}}
\label{barP}
\end{figure}

\end{document}